\documentclass[preprint,aps]{revtex4}
\usepackage{graphicx}  % uncomment this if your want to insert figures
\usepackage{bm}
\usepackage{epsfig}
\usepackage{color}
\newcommand{\be}{
\begin{equation}
}
\newcommand{\ee}{
\end{equation}
}
\newcommand{\ba}{
\begin{eqnarray}
}
\newcommand{\ea}{
\end{eqnarray}
}
\newcommand{\vc}[1]{
{\bf{#1}}
}
\newcommand{\pard}[2]{
{\frac{\partial #1}{\partial #2}}
}
\newcommand{\totd}[2]{
{\frac{{\rm{d}} #1}{{\rm{d}} #2}}
}
\newcommand{\df}{
{\rm{d}}
}
\newcommand{\gav}[1]{
\langle #1 \rangle
}
\newcommand{\intl}{
\int \limits
}

\newcommand{\graph}[2]{
\newpage
\begin{figure}[hhh]
\includegraphics[width=3in,angle=0]{#1.eps}~
\caption{(Color online) #2}
\label{#1} 
\end{figure}
}
\newcommand{\graphnew}[2]{
\newpage
\begin{figure}[hhh]
\includegraphics[width=3in,angle=0]{#1.eps}~
\caption{(Color online) #2}
\label{#1} 
\end{figure}
}
\begin{document}
%%%%%%%%%%%%%%%%%%%%%%%%%%%%%%%%%%%%%%%%%%%%%%%%%%%%%%%%
% The title, only the first letter capitalized; if you want to split it in
% two or more lines, put a \\ macro at each line break
% example: 
%   \title{Title: first line\\ second line}
%
\title{Global gyrokinetic particle-in-cell simulations of internal
  kink instabilities}
\author{Alexey Mishchenko\footnote[1]{alexey.mishchenko@ipp.mpg.de}}
\affiliation{Max-Planck-Institut f\"ur Plasmaphysik,
         EURATOM-Association, 
         D-17491 Greifswald, Germany}
\author{Alessandro Zocco$^{1,2}$}
\affiliation{1) Euratom/CCFE Fusion Association, Culham Science Centre,
  Abingdon, Oxon, UK, OX14 3DB \\ 
  2) Rudolf Peierls Centre for Theoretical
  Physics, 1 Keble Road, Oxford, UK, OX1 3NP}
%
 % \author{Trach-Minh Tran}
%  %
%  \affiliation{\`Ecole Polytechnique F\`ed\`erale de Lausanne, Centre de
%    Recherches en Physique des Plasmas, EURATOM-Association, CH-1015 Lausanne,
%    Switzerland}
 %
%  \author{Roman Hatzky}
%  %
%  \affiliation{High Level Support Team (HLST),
%          Max-Planck-Institut f\"ur Plasmaphysik,
%          EURATOM-Association,
%          D-85748 Garching, Germany}
% %
% \author{Axel K\"onies}
% %
% \affiliation{Max-Planck-Institut f\"ur Plasmaphysik,
%          EURATOM-Association,
%          D-17491 Greifswald, Germany\quad}
%
\date{\today}
%
%%%%%%%%%%%%%%%%%%%%%%%%%%%%%%%%%%%%%%%%%%%%%%%%%%%%%%%%%%%%%%%%%
%
\begin{abstract}
Internal kink instabilities have been studied in straight tokamak geometry
employing an electromagnetic gyrokinetic particle-in-cell (PIC)
code. The ideal-MHD internal kink mode and the collisionless $m = 1$  tearing
mode have been successfully simulated with the PIC code. Diamagnetic
effects on the internal kink modes have also been investigated. 
\end{abstract}
%~~~~~~~~~~~~~~~~~~~~~~~~~~~~~~~
%
\maketitle
%
%%%%%%%%%%%%%%%%%%%%%%%%%%%%%%%%%%%%%%%%%%%%%%%%%%%%%%%%%%%%%%%%%%
%
\newpage
%
%%%%%%%%%%%%%%%%%%%%%%%%%%%%%%%%%%%%%%%%%%%%%%%%%%%%%%%%
% Write the text starting from here and using the usual
% LaTeX commands.
%
%=================================================================
%

%
\section{Introduction}

The internal kink instability is a fluid fixed-boundary mode which
can, however, be strongly affected by kinetic effects. 
It occurs if the
safety factor $q(r_c) = 1$ at some position $r_c$ inside the plasma
column \cite{Newcomb_60,Shafranov_70,Rosenbluth73}. The fluid internal kink mode is always unstable in screw
pinch geometry, but it can be partially stabilized by toroidal
corrections \cite{Bussac75}. Nevertheless, it plays an important role
in tokamak operation (in particular during ``sawtooth''
\cite{sawteeth_ST} and ``fishbone'' \cite{fishbones_PDX} activity).   
In the fluid picture, the internal kink mode is destabilized by the gradients of the
parallel current and of the plasma pressure. Physically,
the kink instability is a tilt and a shift of the plasma column. It
can be unstable because the field line bending vanishes at the position
$r_c$ and fails to compensate the magnetic force associated with the
gradient of the ambient parallel current and the gradient of the
pressure (directly related to the diamagnetic current). The uncompensated perturbed 
magnetic energy is then set free in 
the narrow ``inertial'' (or ``resonant'') layer around
$r_c$ and is converted to kinetic energy of poloidally rotating plasma (and
drives reconnection when non-ideal effects are important). 
A kinetic description may be needed in order to address the physics of
the inertial (resonant) layers. 

The internal kink instability has been known in plasma physics
research from the beginning of the fusion programme. Already in 1973,
it was considered as a plausible candidate to explain the $q(0) > 1$ tokamak
stabilitiy criterion. It has been
hypothesized \cite{Rosenbluth73} that the internal kink mode can be
involved in the disruptive instabilities.  In
Ref.~\cite{Rosenbluth73}, an ideal Magnetohydrodynamical (MHD) theory
of the internal kink modes was developed. A nonlinear kinked
neighboring equilibrium was found. However, the resulting
nonlinear ideal-MHD amplitudes were too small to explain the disruptive
phenomena in tokamaks although such qualitative properties as
the negative voltage spikes and inward shifts in major radius produced by
the mode evolution agreed with the experimental observations. 
The next time the internal kink modes received increased attention
was in 1975 when B.~Kadomtsev proposed his model \cite{Kadomtsev_sawteeth} for the
sawtooth phenomena, discovered experimentally one year before
\cite{sawteeth_ST}. In this model, the growth of the magnetic island 
associated with the internal kink mode (poloidal mode number $m=1$) plays a crucial role. Clearly,
such a process is beyond the range of the ideal MHD theory considered
in \cite{Rosenbluth73}. 

This has motivated the development of theoretical
descriptions of the internal kink mode which include various non-ideal
effects. Thus, in Ref.~\cite{Ara77} the internal kink mode has been
considered within a two-fluid model including %the electrical
resistivity, ion-ion collisions and diamagnetic effects (which
were found to be stabilizing). 
Further development focused on a more detailed description of the
resonant layer around the $q = 1$ magnetic surface. 
In Ref.~\cite{Drake78}, the kinetic theory of the $m=1$ internal
instability was considered (kinetic electrons and fluid
ions). Similarly to Ref.~\cite{Ara77}, a diamagnetic stabilization of
the internal kink mode was found. In
Ref.~\cite{BasuCoppi80}, the role of collisionless reconnection
(driven by the electron inertia) on the evolution of the internal kink
mode was studied in the regime $\delta_e \gg \rho_i$ (here
$\delta_e$ is the electron skin depth and $\rho_i$ is the ion
gyroradius), neglecting ion Finite Larmor Radius (FLR)
effects. Such a regime corresponds to very low values of plasma
pressure $\beta \le m_e/m_i$. It has been pointed out, however, that an accurate
treatment of the ion orbits (in terms of their gyro-average) is needed in order to
consider realistic regimes with $\delta_e \le \rho_i$. A technical
difficulty arising here is due to the non-local character of the
gyrokinetic polarization density which results in an integral
quasineutrality equation (see also \cite{Antonsen81}). This problem has been addressed in
Ref.~\cite{Cowley86} in the context of the tearing mode problem. The
integral equation resulting from the non-local ion response has been
solved by an iterative method (assuming the poloidal beta to be
small).  
In Ref.~\cite{PegoraroSchep89}, an alternative approach to this
problem has been suggested based on the Fourier expansion in the
resonant layer with respect to the radial variable. This approach
emphasizes the singular layer but retains the global nature of the mode
through the boundary conditions. An interpolation
formula has been used in order to treat the gyro-averages (it fails
when the density or temperature gradients are too large). 
%The electron inertia has been neglected in Ref.~\cite{PegoraroSchep89}. 
%
In 1991, Porcelli combined the approach of Ref.~\cite{PegoraroSchep89}
concerning the ions with electron inertia effects assuming,
however, an adiabatic or isothermal electron temperature response
(depending on the parameter regime). 
 In Ref.~\cite{Porcelli91}, he 
 found the so-called $m=1$ collisionless tearing mode which dominates
 the ideal-MHD version of the internal kink 
 mode when the ideal drive is sufficiently small (basically, when the
 ion gyroradius %$\rho_i$ 
 exceeds the ideal-MHD inertial length). 
  This development has been used in the formulation of the revised
  sawtooth model \cite{Porcelli_sawteeth} by Porcelli~et~al in 1996 (this
  model takes into account the complex dynamics of the resonant layer
  which includes the ion FLR and reconnection effects).

Despite the great progress achieved in the analytical understanding of
internal kink modes, it is still an area of active
research. Thus, very recently a unified theory of the internal kink
and tearing modes has been developed in Ref.~\cite{Zocco2012}. This
theory provides an accurate treatment of the ion orbits and the
electron temperature response without assuming it to be adiabatic or isothermal.  

On the numerical side, there is a vast literature on simulations of 
internal kink modes using the fluid approach. For example,
Ref.~\cite{Martynov05} analyzes the ideal stability of the internal kink
mode in shaped tokamak plasmas. Recent full-MHD simulations
\cite{Halpern11_PPCF,Halpern11_PoP} studied various nonlinear regimes
of the internal kink mode evolution and included some
two-fluid (e.~g.~diamagnetic) effects. It is however desirable to
have a first principles approach to the internal kink mode since the
kinetic effects can be important in the resonant layer, and two-fluid
models are often derived under a number of assumptions that may be too
simplistic. Such an attempt was undertaken for the first time with a gyrokinetic
\cite{Brizard:nonlin_gyrokin_finite_beta} particle-in-cell \cite{Lee} 
code in 1995. In Ref.~\cite{SydoraNaitou95}, both the linear and nonlinear
evolution of the internal kink mode were simulated, targeting the
sawtooth collapse on a fast time scale. However, these early simulations
operated at extremely small $\beta \le m_e/m_i$ and neglected ion
FLR effects. This work has recently been continued in
Ref.~\cite{Naitou2009} (also at a very small plasma beta). In addition
to particle-in-cell simulations, the internal kink mode has been 
considered by employing an eigenvalue approach to the solution of the
gyrokinetic equations \cite{Qin:gyrokin_perp_dynamic,Lauber_phd}.  

In this paper, we study the linear evolution of the internal kink
mode in straight-tokamak geometry employing the global 
gyrokinetic particle-in-cell method. Realistic values of plasma $\beta
\gg m_e/m_i$ are used. The objective of this paper is two-fold. First, we study how
the kink mode properties change depending on the relation between the
ion gyroradius and the MHD inertial length. Second, we 
explore, for the first time, what could be the limitations of the
gyrokinetic PIC approach to this problem at realistic values of plasma
beta. 

% Second, we want to
% consider the subtleties of the gyrokinetic particle-in-cell approach
% to the kink mode problem since such simulations have never been
% undertaken for the realistic values of plasma beta. 

The structure of the paper is as follows. In Sec.~\ref{theory}, the
basic equations and the discretization procedure are
discussed. Simulations in the fluid and kinetic regimes are presented in
Sec.~\ref{simulations}. Finally, our conclusions are summarized in Sec.~\ref{conclusions}.
\section{Basic equations and numerical approach} \label{theory}
We use the linear two-dimensional $\delta f$ PIC-code GYGLES
\cite{Mishchenko_tae,Mishchenko_fast,Mishchenko_bulk}.
The code is electromagnetic and treats all
particle species (ions and electrons) kinetically. It solves the gyrokinetic Vlasov-Maxwell system of equations
(in the $p_{\|}$-formulation, see
Ref.~\cite{Brizard:nonlin_gyrokin_finite_beta} for details). The
distribution function is split into a 
background part and a small time-dependent perturbation $f_s = F_{0s} + \delta f_s$ (the
index $s = i,e$ is used for the particle species). The background
ion distribution function is taken to be a Maxwellian. The background
electron distribution function is a shifted Maxwellian (to account for the
equilibrium parallel current):
\be
F_{0e} = n_0 \left(\frac{m}{2 \pi T_e}\right)^{3/2}
\exp\left[{}- \frac{m_e (p_{\|} - u)^2}{2 T_e}\right]
\exp\left[{}- \frac{m_e v_{\perp}^2}{2 T_e}\right]
\ee
Here $u = {}- j_{\|}^{(0)}/(e n_0)$ and $\mu_0 \, j_{\|}^{(0)} =
(\vc{B}/B) \cdot (\nabla \times \vc{B})$ with $\vc{B}$ the
equilibrium magnetic field.

Assuming the amplitude of the field perturbation to be small (this
implies $\delta f_s/F_{0s} \ll 1$), 
%since $\delta f_s$ is a function of the perturbed field), 
the first-order perturbed distribution function
%$\delta f_s$ 
can be found from the linearized Vlasov equation:
\be
\label{vlasov_lin}
\pard{\delta f_{s}}{t} + \dot{\vc{R}}^{(0)} \cdot \pard{\delta f_{s}}{\vc{R}} +
\dot{p}_{\|}^{(0)} \pard{\delta f_{s}}{p_{\|}} = 
{}- \dot{\vc{R}}^{(1)} \cdot \pard{F_{0s}}{\vc{R}} - 
\dot{p}_{\|}^{(1)} \pard{F_{0s}}{p_{\|}} .
\ee
Here, $[\dot{\vc{R}}^{(0)}, \dot{p}_{\|}^{(0)}]$ correspond to the
unperturbed gyrocenter position and parallel velocity,  and 
$[\dot{\vc{R}}^{(1)}, \dot{p}_{\|}^{(1)}]$ are the perturbations 
of the particle trajectories proportional to the electromagnetic field
fluctuations [shown in Eqs.~(\ref{dotR_p0})-(\ref{dotp_p1}) below]. 
The perturbed part of the distribution function is discretized with
markers (see Ref.~\cite{Fivaz_1998} for details): 
%according to the Klimontovich representation: 
\be
\label{pic0}
\delta f_s(\vc{R},p_{\|},\mu,t) = \sum_{\nu=1}^{N_p} 
w_{s\nu}(t)  %\frac{w_{s\nu}(t)}{2\pi J}
\delta(\vc{R} - \vc{R}_{\nu}) \delta(p_{\|} - p_{\|\nu}) 
\delta(\mu - \mu_{\nu}) \ ,
\ee
where $N_p$ is the number of markers, $(\vc{R}_{\nu},p_{\|\nu},\mu_{\nu})$
are the marker phase space coordinates and $w_{s\nu}$ is the weight of a
marker.
The equations of motion are \cite{Brizard:nonlin_gyrokin_finite_beta}
\ba
\label{dotR_p0}
\vc{\dot{R}}^{(0)} &=& p_{\|} \vc{b}^{*} +
\frac{1}{q B_{\|}^{*}} \vc{b} \times \mu \nabla B  \\
\vc{\dot{R}}^{(1)} &=& {}- \frac{q}{m} \gav{A_{\|}} \vc{b}^{*} +
\frac{1}{B_{\|}^{*}} \vc{b} \times \left( \nabla \gav{\phi}
- p_{\|} \nabla \gav{A_{\|}} \right)   \\
\dot{p}_{\|}^{(0)} &=& {}-\frac{\mu \nabla B}{m} 
\cdot \vc{b}^{*} \\
\label{dotp_p1}
\dot{p}_{\|}^{(1)} &=& {}-\frac{q}{m}  \left( \nabla
\gav{\phi}  - p_{\|} \nabla \gav{A_{\|}} \right) 
\cdot \vc{b}^{*}
\ea
with $\phi$ and $A_{\|}$ being the perturbed electrostatic and magnetic
potentials, $\mu$ the magnetic moment, $m$ the mass of the particle, 
%The equilibrium magnetic field enters through the quantities 
$B_{\|}^{*} = \vc{b} \cdot \nabla \times \vc{A}^{*}$,  
$\vc{b}^{*} = \nabla \times \vc{A}^{*} / B_{\|}^{*}$,  
$\vc{A}^{*} = \vc{A} + (m p_{\|}/q) \vc{b}$  
the so-called  modified vector potential, $\vc{A}$ the magnetic
potential corresponding to the equilibrium magnetic field $\vc{B} = \nabla
\times \vc{A}$ and $\vc{b} = \vc{B}/B$ the unit vector in the direction of the
equilibrium magnetic field.
The gyro-averaged potentials are defined as usual:
\be
\label{gyroav}
\gav{\phi} = \oint \frac{\df \theta}{2 \pi} \phi (\vc{R} + \bm{\rho}) \;,\;\;
\gav{A_{\|}} = \oint \frac{\df \theta}{2 \pi} A_{\|} (\vc{R} +
\bm{\rho}) \ ,
\ee
where $\bm{\rho}$ is the gyroradius of the particle and $\theta$ is the
gyro-phase. Numerically, the gyro-averages are computed sampling
a sufficient number of the gyro-points on the gyro-ring around the gyrocenter position of
the marker \cite{Lee,Mishchenko3}. 

The perturbed electrostatic and magnetic potentials are found
self-consistently from the gyrokinetic quasineutrality equation and
parallel Amp\'ere's law \cite{Brizard:review}: 
\ba
\label{qasi}
&&{} \int \frac{q_i F_{0i}}{T_i} \, (\phi - \gav{\phi}) \, \delta(\vc{R} +
\bm{\rho} - \vc{x}) \, \df^6 Z  
 = \delta n_i - \delta n_e \\
\label{amp}
&&{} \left( \frac{\beta_i}{\rho_{i}^{2}} +
\frac{\beta_e}{\rho_{e}^{2}} - 
\nabla_{\perp}^{2} \right) A_{\|} = \mu_{0} 
\left( \delta j_{\|i} + \delta j_{\|e} \right) \ ,
\ea
where $\delta n_{s} = \int \df^{6}Z \, \delta f_{s} \, \delta(\vc{R} +
\bm{\rho} - \vc{x})$ is the gyrocenter density,  
$\delta j_{\|s} = q_{s} \int \df^{6}Z \, \delta f_{s} \, p_{\|} \,
\delta(\vc{R} + \bm{\rho} - \vc{x})$ is the gyrocenter current, $q_s$ is the
charge of the particle, $\df^{6}Z = B_{\|}^{*} \, \df \vc{R} \, \df p_{\|} \,
\df \mu \, \df \theta$ is the phase-space volume, 
$\rho_{s} = \sqrt{m_{s}T_{s}}/(eB)$ is the thermal gyroradius and 
$\beta_s = \mu_{0} n_{0} T_{s}/B_{0}^{2}$ is the plasma beta
corresponding to a particular species. Note that the polarization
density in Eq.~(\ref{qasi}) is given by an integral operator and
includes a non-local effect of the ion gyro-orbit.

We have found, however, that the computations can be simplified 
by replacing the quasineutrality condition, Eq.~(\ref{qasi}), with 
\be
f_c(r) \int \frac{q_i F_{0i}}{T_i} \, (\phi - \gav{\phi}) \, \delta(\vc{R} +
\bm{\rho} - \vc{x}) \, \df^6 Z  - [1 - f_c(r)] \; \nabla \cdot \left( \frac{q_i n_{0}}{T_{i}} \rho_{i}^{2}
  \nabla_{\perp} \phi \right)  = \delta n_i - \delta n_e
\ee
Here, 
\be
f_c(r) = \exp\left[ {}- \left( \frac{r - r_c}{\Delta r_c} \right)^{p_c} \right]
\ee
with $r_c$ being the position of the singular layer. This
representation implies that the exact expression for the polarization
density is used around the resonant position $q(r_c) = 1$ (where the
relevant radial scale can be smaller than the ion gyroradius), whereas the long-wavelength
approximation is used outside the resonant layer (which is justified since the kink mode
structure is global in the ideal region).  
 The effect of this representation is to replace (in the ideal region
 only!) the kinetic Alfv\'en waves by shear Alfv\'en waves, which
 have more favourable (for numerics) properties experiencing more
 physical damping on small radial scales.  
In our simulations,
we have used $\Delta r_c = 0.2 r_a$ and $p_c = 8$. For this choice of
the parameters, $\rho_i \ll \Delta r_c$ in all cases considered. 

The electrostatic and magnetic potentials are discretized with the
finite-element method (Ritz-Galerkin scheme): %according to the expressions:
\be
\label{fin_el_discr}
\phi(\vc{x}) = \sum_{l=1}^{N_s} \phi_l \Lambda_l(\vc{x}) \ , \;\;\;
A_{\|}(\vc{x}) = \sum_{l=1}^{N_s} a_l \Lambda_l(\vc{x}) \ ,
\ee
where $\Lambda_l(\vc{x})$ are finite elements (tensor products of B
splines \cite{de_Boor,Hoellig}), $N_s$ is the total
number of the finite elements, $\phi_l$ and $a_l$ are the spline
coefficients. 
The numerical treatment of the nonlocal gyrokinetic polarization density has been
described in Ref.~\cite{Mishchenko3}. Homogeneous Dirichlet boundary
conditions are applied for $\phi$ and $A_{\|}$ both on the axis and at the plasma edge.
Further details of the numerical discretization can
be found in Refs.~\cite{Fivaz_1998,Mishchenko1,Mishchenko2,Mishchenko3,Hatzky_2007}. 
%
%=================================================================
%
\section{Simulations} \label{simulations}
We consider a straight tokamak (a screw pinch) with minor radius $r_a$
and ``major radius'' $R_0$ (the pinch is a topological torus with the
length $L = 2 \pi R_0$ and periodic boundary conditions along the axis). The plasma
consists of hydrogen ions and electrons (with a realistic mass ratio). The safety
factor is given by the expression $q(r) = q_0 + (1 - q_0)
(r/r_c)^{p_q}$ (with $q_0 = 0.6$ and $p_q = 1$ in our
simulations). The background magnetic field is determined by the MHD 
pressure balance condition:
\be
\totd{}{r} \left( p + \frac{B_z^2 + B_{\theta}^2}{2 \mu_0} \right) +
\frac{B_{\theta}}{\mu_0 r} = 0 \ , \;\;\; B_{\theta} = 
\frac{r}{q(r) R_0} \, B_z \; .
\ee
This equation is solved with the ``initial condition'' $B_z(r=0) =
B_0$. It can be seen that $B_z(r) \approx B_0$ often gives a good
approximation for the equilibrium magnetic field (when the effect of
the plasma pressure is small).

The plasma temperature and density profiles are given by the expressions:
\ba
\label{nT0_eq}
n_0(s) = {N}_0 \exp\left[ {}- \frac{\Delta_{{\rm n}}}{L_{{\rm n}}} \,
{\rm tanh}\left(\frac{s - s_{{\rm n}}}{\Delta_{{\rm n}}}\right)
\right] \ , \;\;\;
%\label{T0_eq}
T_{s}(s) = T_{0s} \exp\left[ {}- \frac{\Delta_{{\rm T}}}{L_{{\rm T_s}}} \,
{\rm tanh}\left(\frac{s - s_{{\rm T}}}{\Delta_{{\rm T}}}\right) \right] 
\ea
where $s = r/r_a$ with $r_a$ the radius of the pinch. The shape of these profiles can
be flexibly tailored by adjusting the parameters $\Delta_{{\rm T}}$ and 
  $\Delta_{{\rm n}}$ (which determine the profile width), $s_{{\rm T}}$ and 
  $s_{{\rm n}}$ (position of the maximal gradient), $L_{{\rm T_s}}$
  and $L_{{\rm n}}$ (gradient lengths).
% Other parameters are $s_{{\rm T}} = 0.5$ the position of the maximal $\kappa_{{\rm T}} =
% |\nabla T_i| / T_i$, $T_{0}$ the bulk plasma temperature at $s = s_{{\rm T}}$,
% $\Delta_{{\rm T}} = 0.2$ the ``width'' of the
% bulk plasma temperature profile, and $L_{{\rm T}} = 0.3$ the ``length'' of the
% bulk plasma temperature profile (this parameter determines how large the
% temperature gradient is).

The numerical resolution needed in the simulations depends on the
physical parameters. It has been observed that the radial resolution
must always be sufficient to resolve the ion gyroradius
(it is required since the Alfv\'en continuum must be resolved). For
cases when the electron inertia is of importance, the electron
skin depth must also be resolved. In the ``poloidal'' direction, 
$N_{\theta} = 4$ splines have been found sufficient since only
one poloidal mode is kept in the linear straight-tokamak
simulations. The marker resolution is kept on the level of $700-1000$
markers per grid cell. 

% are as follows:
% the number of ion markers $N_i = 8 \times 10^6$, the number of electron markers
% $N_e = 8 \times 10^6$, the
% number of radial grid points $n_r = 800$, the time step $\Delta t = 2 \times 10^{-7}$~{\rm s}, the
% number of iterations in order to solve the cancellation problem
% $N_{\rm iter} = 2$. Convergence studies have shown that these parameters
% correspond to well converged simulations.
%
%~~~~~~~~~~~~~~~~~~~~~~~~~~~~~~~~~~~~~~~~~~~~~~~~~~~~~~~~~~~~~~~~~
%
\subsection{Internal kink mode: ``MHD regime''} \label{MHD_regime}
In this subsection, we consider the internal kink instability in the MHD
regime. Physically, it implies that the width of the resonant layer
(which in this case coincides with the MHD {\it inertial-layer width} $\lambda_H$) is much larger than the ion
gyroradius $\rho_i$. The inertial-layer width is given by the expression \cite{Rosenbluth73,Ara77}:
\be
\label{lambda_H}
\frac{\lambda_H}{r_c} = {}-\frac{\pi}{[r_c q'(r_c) B_{\theta}(r_c)]^2} \; \intl_0^{r_c}
g_1(r) \df r 
\ee
with
\be
\label{g1}
g_1(r) = \frac{(m^2 - 1) r (\vc{k}\cdot\vc{B})^2}{m^2 + k_z^2 r^2} +
\frac{k_z^2 r^2}{m^2 + k_z^2 r^2} \left( 2 \mu_0 \totd{p}{r} + r
  (\vc{k}\cdot\vc{B})^2 + \frac{2}{r} \, \frac{k_z^2 r^2 B_z^2 - m^2 B_{\theta}^2}{m^2 + k_z^2 r^2}
\right) \; .
\ee
Here, $q' = dq/dr$, $k_z = n/R_0$ and $\vc{k} \cdot \vc{B} = B_z \, (
m - q n ) / (q R_0)$. For the internal kink mode, the poloidal mode number
$m=1$ and the toroidal mode number $n=1$.

Analytical ideal-MHD theory \cite{Rosenbluth73} gives the following
expression for the growth rate:
\be
\label{gamma_MHD}
\gamma \tau_A = q'(r_c) \lambda_H \ , \;\;\; \tau_A = R_0 / v_A \; .
\ee
Here, $v_A = B_z/\sqrt{\mu_0 m_i n_0}$ is the Alfv\'en
velocity. Another way (used here) is to solve numerically (with the
shooting method) the ideal-MHD eigenvalue problem: 
\be
\label{MHD_eigen}
\totd{}{r}\left([\mu_0 m_i n_0 \gamma^2 + (\vc{k}\cdot\vc{B})^2] r^3
  \totd{\xi}{r}\right) - g_1(r) \xi = 0
\ee
employing the boundary conditions 
$\xi'(r=0) = 0$ and $\xi(r=r_a) = 0$. Here, $\xi(r)$ is the MHD displacement 
and $\xi' = \df \xi / \df r$. 

We consider a straight tokamak with ``major radius'' $R_0 = 5$~{\rm m},
minor radius $r_a = 1$~{\rm m}, magnetic field $B_0
= 2.5$~{\rm T}, plasma temperature $T_i = T_e = 5$~{\rm keV}, and
plasma density $N_0 = 2 \times 10^{19}~{m^3}$ (which corresponds
to $\beta = 0.0128$ at the temperature chosen). Both the plasma
density and the temperature are 
flat. For these parameters, an unstable internal kink mode exists
(destabilized by the gradient of the ambient parallel current).

We consider a sequence of straight-tokamak equilibria
corresponding to different locations of the rational flux surface $r_c$.
In Fig.~\ref{kink-b10q0.6_qc_scales},
the ideal-MHD inertial-layer width $\lambda_H$ is shown as a function
of $r_c$. One sees that $\lambda_H$ exceeds both the ion gyroradius
$\rho_i = 2.8 \times 10^{-3}$~{\rm m} and the electron skin depth $\delta_e =
\rho_e/\sqrt{\beta_e} = 1.7 \times 10^{-3}$~{\rm m} for all equilibria considered.  

In Fig.~\ref{kink1-b10q0.6_qc_ti_g}, the internal kink mode growth rate resulting
  from the gyrokinetic PIC simulations is compared to the results of
  the ideal-MHD eigenvalue 
  calculations [the MHD eigenvalue problem Eq.~(\ref{MHD_eigen}) has
  been numerically solved 
  using the shooting method]. One sees that the mode is more unstable
  for larger $r_c$, i.~e.~when a larger plasma column is involved in the
  instability [this can also be seen formally from Eq.~(\ref{lambda_H})]. 
  The quantitative agreement between the
  ideal MHD and the gyrokinetic simulations is very good when the ion
  temperature is 
  small $T_i = 200$~{\rm keV} (implying ion gyroradius small). When
  the ion gyroradius increases, the kink mode becomes less unstable
  (although the quialitative dependence on $r_c$ remains 
  the same). The mode is still an MHD-like internal kink
  instability but this will change when $k_{\perp} \rho_i > 1$ with
  $k_{\perp} \sim 1/\lambda_H$ (see Sec.~\ref{FLR_regime}).   
One can quantify the ion-FLR effects considering a sequence of internal
kink modes keeping all the parameters constant except the plasma
temperature ($T_i = T_e$).  Note that the ideal-MHD growth rate does
not depend on the temperature (for a flat profile such as used in
these simulations). In contrast, the gyrokinetic simulations show that
the kink mode growth rate decreases with the plasma temperature
(see Fig.~\ref{kink1-b10q0.6_qc0.5_ti-g}). This FLR-stabilization
effect can be quite substantial when the ion  temperature is large enough. 
Note that a similar ion-FLR stabilization has already been observed for
other MHD modes (see the gyrokinetic simulations of the
Toroidal Alfv\'en Eigenmodes in
Ref.~\cite{Mishchenko_fast}). The stabilization is 
caused by the gyro-averaging operation acting on the perturbed
electromagnetic field. Clearly, this effect is absent in the
ideal MHD description and can be found only with a kinetic treatment. 

In Fig.~\ref{kink1-b10q0.6_qc0.5_ti200_s}, we compare the ideal-MHD
eigenmode structure with the radial pattern found in the gyrokinetic
simulations. An excellent agreement is found. The ideal-MHD
displacement $\xi \sim \phi/r$ corresponding to
Fig.~\ref{kink1-b10q0.6_qc0.5_ti200_s} has the well-known top-hat
structure. The poloidal fluid velocity $v_{\theta}
\sim \partial \phi / \partial r$ is clearly strongly increased at the rational
flux surface, as must be the case for the internal kink mode. 

In Fig.~\ref{kink-b10q0.6_qc0.7_t}, the internal kink mode evolution is
shown (here, the rational flux surface is located
  at $r_c = 0.7$; the plasma temperature $T_i = T_e = 5$~{\rm
    keV}). One sees how the initial perturbation (which has had a Gaussian
  shape in the radial direction) decays in the continuum of the shear
  Alfv\'en waves and is then re-organized as an unstable internal kink
  mode. The ion gyroradius must be resolved in order to correctly reproduce
  the continuum-decay phase of the mode evolution. The
  frequency of this mode is zero (this is a well-known property of the
  ideal MHD modes, related to the Hermitian symmetry of the underlying 
  equations; this property can also be reproduced in the gyrokinetic
  simulations).  %when the diamagnetic effects are weak).

So far, we have considered the current-driven internal kink mode (the
plasma pressure was chosen to be flat). Now, let us include the
pressure destabilization in our simulations. In
Fig.~\ref{kinkB5-q0.6qc0.7_kn-g}, the kink-mode growth rate is plotted
as a function of the plasma density gradient. For the density profile,
we choose $s_n = 0.5$, $\Delta_n = 0.2$ [see
Eq.~(\ref{nT0_eq})]. The plasma temperature $T_i = T_e = 3$~{\rm keV}
is taken to be flat. The resonant flux surface has
been located at $r_c = 0.7$. Two cases, one with a large magnetic field $B =
5$~{\rm T} and the other with a moderate magnetic field $B = 2.5$~{\rm T}, were
considered [keeping $\beta(s_n) = 0.00773$ constant]. One sees from
Eq.~(\ref{gamma_MHD}) that the
ideal-MHD result does not depend on the magnetic field
strength. Indeed, the gyrokinetic simulations reproduce this
property. The agreement between the gyrokinetic
simulations and the MHD result is good for both values of the ambient
magnetic field. The kink-mode growth rate increases with the density
(pressure) gradient, as expected. 

Summarizing, we have considered the internal kink mode in the regime
$\rho_i < \lambda_H$ (ions in the resonant layer are magnetized). In this
regime, the properties of the kink mode agree well with the ideal MHD
expectations (thus providing a benchmark for our simulations). The
only effect found beyond ideal MHD, is an ion-FLR stabilization
of the internal kink mode (which however can be quite substantial). 
%
%~~~~~~~~~~~~~~~~~~~~~~~~~~~~~~~~~~~~~~~~~~~~~~~~~~~~~~~~~~~~~~~~~
%
\subsection{Internal kink mode: ``FLR regime''}   \label{FLR_regime}
We now consider the case when the thermal ion gyroradius exceeds the
ideal-MHD inertial width $\rho_i \gg \lambda_H$. The parameters here are as
follows. The straight tokamak has the minor radius $a = 0.5$~{\rm m}, the
ambient magnetic field $B_0 = 1$~{\rm T}, and the plasma density $n_0
= 1.6 \times 10^{18}~{m^{-3}}$ (it corresponds to $\beta = 0.00644$
when $T_i = T_e = 5$~{\rm keV}). Equilibria with the major radii
$R_0 = 5$~{\rm m} and $R_0 = 20$~{\rm m} have been considered (changing the
major radius one can change the aspect ratio and, consequently, the
inertial-layer width $\lambda_H$ while $\rho_i/r_a$ is kept
fixed). The comparison between the inertial layer width    
$\lambda_H$, the thermal ion gyroradius
$\rho_i$, and the electron skin depth $\delta_e$ is shown for different
aspect ratios in 
Fig.~\ref{kink-tp_b1.0q0.6_qc_ti5e3_scales}.  One sees that the
ion gyroradius is indeed much larger than all other relevant scales,
especially when $R_0 = 20$~{\rm m} (this case corresponds to a
particularly weak, almost vanishing ideal-MHD drive). 

In this regime (ions are demagnetized in the resonant layer),
Ref.~\cite{Porcelli91} predicts the existence of an unstable collisionless
$m=1$ tearing mode. In the case $\lambda_H \rightarrow 0$, the  growth
rate of this mode is given by the expression \cite{Porcelli91}:
\be
\label{Porcelli_kink}
\gamma = \frac{\hat{s}_q(r_c) v_A(r_c)}{R_0} \; \frac{(\delta_e
  \rho_i^2)^{1/3}}{r_c} \ , \;\;\;
v_A = \frac{B_0^2}{\sqrt{\mu_0 m_i n_0}} \ , \;\;\;
\hat{s}_q = \frac{r}{q} \, \totd{q}{r} .
\ee
In Fig.~\ref{kink-b1.0R20.0qc-gs_ti11e3-g}, we plot the growth rate resulting
  from the gyrokinetic PIC simulations (corresponding to the pinch
  with $R_0 = 20$~{\rm m}, i.~e.~very small $\lambda_H$) compared to
  the analytic prediction for the collisionless $m=1$ tearing mode as a function of
  the resonant layer position. In contrast to the ``ideal-MHD''
  regime (see Fig.~\ref{kink1-b10q0.6_qc_ti_g}), the mode is more
  unstable for smaller $r_c$, indicating a change in the underlying
  physical mechanism: reconnection (electron physics) in
  addition to the poloidal plasma rotation (ions). The agreement
  between the gyrokinetic result and the theoretical prediction
  Eq.~(\ref{Porcelli_kink}) is very good. 

In Fig.~\ref{kink_b1.0R20.0qc0.4-gs_ti_g}, we consider the dependence
of the kink-mode growth rate on the ion temperature (ion-FLR
effect). One sees that in the regime of demagnetized ions (inside 
the resonant layer), the mode is further destabilized when the ion
gyroradius increases. This FLR-destabilization is in contrast to the 
MHD-type internal kink mode considered previously, which was
FLR-stabilized (see Fig.~\ref{kink1-b10q0.6_qc0.5_ti-g}).  The
numerical result has been compared with the analytical expression
Eq.~(\ref{Porcelli_kink}). The agreement is again very good. 
For comparison, we plot the growth rate predicted by the ideal MHD
theory, which is more than two orders of magnitude smaller. 
%than the growth rate of the kinetic mode. 
It indicates that the kinetic effects can dominate the kink mode
physics, certainly around ideal marginal stability. 

Finally, we consider the effect of the electron temperature gradient on
the collisionless $m=1$ tearing mode. We employ a temperature
profile with $s_{{\rm T}} = r_c/r_a =
0.5$ (the position of the maximal gradient coincides with that of the
resonant flux surface) and $\Delta_s = 0.2$ [see
Eq.~(\ref{nT0_eq})]. The ion temperature and the plasma density are
taken to be flat. As a consequence of the diamagnetic effect
associated with the finite electron temperature gradient, the mode
acquires a finite frequency and becomes a drift-tearing
instability (see e.~g.~\cite{Cowley86,Zocco2012}). The frequency resulting
from the gyrokinetic PIC simulations is plotted in Fig.~\ref{kink-b1.0R5q0.6_qc0.5_kTe-w} as a function of $L_{Te}$ (electron
temperature gradient length) for two different values of the
major radius: $R_0 = 5$~{\rm m} and $R_0 = 10$~{\rm m} (recall that the
ideal-MHD drive and, consequently, 
$\lambda_H$ {\it decrease} with the aspect ratio). One sees that the
frequency is the same for both values of $R_0$. Hence, it is set
exclusively by the diamagnetic frequency at the position of the
resonant flux surface $\omega_{{\rm T}e} = 1/(B r_c) \, \df T_e/\df
r$. In fact, the frequency of the collisionless $m=1$ drift-tearing
mode appears to satisfy $\omega \approx 0.4 \, \omega_{{\rm T}e}$
(cf.~Ref.~\cite{Cowley86}). 
%which is in agreement with Eq.~(87) of Ref.~\cite{Cowley86}. 

The growth rate as a function of $L_{Te}$ (electron
temperature gradient length) is plotted in
Fig.~\ref{kink-b1.0R5q0.6_qc0.5_kTe-g}. As expected
\cite{Cowley86,Zocco2012}, the electron temperature gradient
stabilizes the collisionless drift-tearing $m=1$ mode. Physically, it
has been shown in Ref.~\cite{Zocco2012} that the drift-tearing mode
can couple to a stable Kinetic Alfv\'en Wave (KAW) which leads to a
stabilization of the mode. Another way to explain this stabilization
effect has been elaborated in 
Ref.~\cite{Breizman_fishbones} (in the fishbone context). In short, the
finite-frequency mode can interact with the Alfv\'en continuum (or,
alternatively, with the KAWs as in \cite{Zocco2012}) and
thus undergo continuum damping (as is well-known from the
context of Toroidal Alfv\'en Eigenmodes
\cite{Zonca_Chen_92,Berk_conti_92, Rosenbluth_conti_92}). Of course,
the frequency of the drift-tearing $m=1$ mode is quite small, so that
the continuum damping acts only very close to the resonant flux
surface (the surface of vanishing $k_{\|}$) where the condition $\omega^2 = 
k_{\|}^2(r_A) v_A^2(r_A)$ can be satisfied. The frequency of the $m=1$
drift-tearing mode increases with the electron temperature gradient
which makes the continuum damping more efficient at smaller $L_{Te}$
-- to the point of a complete stabilization. 
Note that in Fig.~\ref{kink-b1.0R5q0.6_qc0.5_kTe-g} the stabilization
appears to be stronger for the case with the 
larger major radius $R_0 = 10$~{\rm m}. In this case, the absolute value of
the growth rate would be smaller compared with the configuration with
$R_0 = 5$~{\rm m}, even for 
the flat temperature profiles, because the growth rate inversely scales with the
Alfv\'en time $\tau_A \sim R_0$. Thus, the continuum damping, 
which is determined by the mode frequency (set by the electron
temperature gradient and thus equal for both values of $R_0$), corresponds to a
larger fraction of the {\it smaller growth rate} when $R_0 = 10$~{\rm m}.  

In Fig.~\ref{kink-b1.0R10q0.6_qc0.5_kTe_sm}, the radial mode structure
is shown. One sees that in addition to the conventional internal kink
eigenmode, a complicated fine-scale structure appears at the resonant
flux surface. This structure is caused by the continuum damping of the
instability. One can see that it becomes more pronounced when the
electron temperature gradient increases (see
Fig.~\ref{kink-b1.0R10q0.6_qc0.5_kTe_sm_small}). 
We can estimate the position of the shear Alfv\'en resonances using
the expression: 
\be
r_A = r_c \pm \frac{\omega \tau_A}{q'(r_c)}  \ , \;\;\; \tau_A = R_0 / v_A
\ee
For the safety factor profile chosen, $q'(r_c) =
0.8$. Fig.~\ref{kink-b1.0R10q0.6_qc0.5_kTe_sm_small} indicates that
the fine-scale structure developed could be understood as Kinetic
Alfv\'en Waves (recall that we use a non-local expression for the
polarization density) resonantly excited by the collisionless $m=1$
tearing mode in the positions approximately satisfying the shear Alfv\'en wave
resonance condition $\omega^2 \approx k_{\|}^2(r_A) v_A^2(r_A)$ (the Kinetic Alfv\'en
Wave resonance would be more precise).
% to use  condition $\omega \approx \pm k_{\perp} \rho_i \; k_{\|}
% v_A$, see e.~g.~\cite{Schekochihin2009}). 
%since the sub-gyro radial scales are involved;
%this would shift the resonance position closer to $r = r_c$ when $k_{\perp}
%\rho_i > 1$). 
The drift-tearing  (kinetic-kink) mode is damped because of its 
coupling to the Kinetic Alfv\'en Waves (a ``continuum'') at the
resonant positions. A finite frequency of the tearing mode is required
for this process to function. This frequency is provided by the
diamagnetic effect due to a finite electron-temperature gradient. 

\section{Conclusion} \label{conclusions}
In this paper we have studied the internal kink modes in a
straight-tokamak geometry using the global gyrokinetic particle-in-cell code
GYGLES. Both electron and ion gyrocenters were treated kinetically,
but collisions were ignored. 
The simulations have shown that the kink mode properties depend strongly
on the ratio between the ideal-MHD inertial-layer scale and the
gyroradius. In the ``MHD regime'', the kink mode becomes more unstable
if the rational flux surface ``moves'' (during the parameter scan) outwards. In the ``FLR regime'',
however, the kinetic-kink mode (the ``collisionless $m=1$ tearing mode''
\cite{Porcelli91}) is more unstable for resonant magnetic surfaces
located closer to the magnetic axis ($1/r_c$ dependence of the growth
rate). Similarly, the scaling with respect to the ion temperature is
also opposite: the FLR-stabilization for the MHD-type kink mode has
been observed whereas the ``collisionless $m = 1$ tearing mode'' is
FLR-{\it destabilized}. One could speculate \cite{Porcelli91} that the most
unstable $m=1,n=1$ instabilities observed in real hot plasmas
often correspond to the ``collisionless $m=1$ tearing'' (kinetic-kink) modes rather than the
classical ideal-MHD internal kink modes since usually the real plasmas are
at the marginal MHD-kink stability boundary so that the ion gyroradius can
easily overcome the ideal inertial length (thus bringing the instability into
the kinetic regime). The kinetic-kink modes, however, can be sensitive to the
shape of plasma profiles and can be stabilized e.~g.~by the electron
temperature gradient (through a combination of the diamagnetic and
continuum-damping effects). 

% Internal kink mode represents an excellent situation where the impact
% of the non-equidistant grid on the simulation results and the code
% performance can be studied. It has been observed that the
% non-equidistant grids (densified at the rational flux surface) can 
% substantially speed up the code convergence with respect to the number
% of the radial knots. This can have important implications for
% non-linear multiple scale simulations.

Looking forward, gyrokinetic simulations of the internal kink modes in tokamak
geometry would be of interest. In these simulations, the effects of the
guiding center orbits in the resonant region, the  kinetic trapped-ion
%(Kruskal-Oberman) 
and fast-particle (e.~g.~fishbone) effects in the
ideal region should be addressed. Also, gyrokinetic simulations of 
interchange instabilities (both in straight and toroidal
geometries) could be performed in the future.

% extension of our simulations to the tokamak
% geometry is planned. The main difficulty in this respect is that the
% internal kink mode is only marginally unstable in tokamak geometry (in
% contrast to the screw pinch). More complicated particle orbits and
% boundary conditions (``lost'' particles) make the cancellation problem
% more difficult to handle in a tokamak (recall that the kink mode numbers
% correspond to small $k_{\perp}$ which is unfavourable for the
% numerical cancellation in Amp\'ere's law \cite{Mishchenko1}). In
% addition, there is a question about the background distribution
% function in tokamak geometry. In the screw pinch, a shifted Maxwellian
% has been used in order to account for the parallel ambient current. In
% tokamaks, at least a solution of the neoclassical stationary problem must be
% applied (or, more accurately, the neoclassical equilibrium must be an
% integral part of the kinetic simulations).
%actually 
%In practice, this may imply a coupling between 
%neoclassical and gyrokinetic codes.
%
%%%%%%%%%%%%%%%%%%%%%%%%%%%%%%%%%%%%%%%%%%%%%%%%%%%%%%%%%%%%%%%%%
\qquad \\
\qquad \\
\qquad \\
{\bf ACKNOWLEDGMENTS}
%%%%%%%%%%%%%%%%%%%%%%%%%%%%%%%%%%%%%%%%%%%%%%%%%%%%%%%%%%%%%%%%%

We acknowledge P.~Helander and A.~ Schekochihin
who have supported this work. R.~Kleiber's help on 
the shooting algorithm is appreciated. We thank J.~Connor for 
carefully reading this manuscript. The simulations have been
performed on the HPC-FF supercomputer ({J\"ulich}, Germany) and the HELIOS
supercomputer (Aomori, Japan) as well as on the local cluster in
Greifswald (H.~Leyh's help is appreciated). Joint work on this paper has
been enabled by the Gyrokinetic Meetings in Vienna and Madrid. We
thank N.~Mauser (Wolfgang Pauli Institute, Vienna) and I.~Calvo (CIEMAT,
Madrid) for organizing these workshops. Support by the Leverhulme
Trust International Academic Network in Magnetised Plasma Turbulence
and Euratom Mobility is also appreciated. 
%
%
%###################################################
%

%=================================================================
%=================================================================
%--------------------------------
%\newpage
%{\bf \LARGE Bibliography}
\bibliographystyle{prsty}

%\bibliography{\BIBL/gyrokin,\BIBL/gygles,\BIBL/dispers,\BIBL/aitg,\BIBL/curv_pinch,\BIBL/kin_mhd}
%-------------------------------------------------

% \begin{figure}[hhh]
% \includegraphics[width=3in,angle=0]{./graph/pinch/kink-b10q0.6_qc_g.eps}~
% \caption{Kink mode in the MHD regime}
% \label{kink-b10q0.6_qc_g} 
% \end{figure}

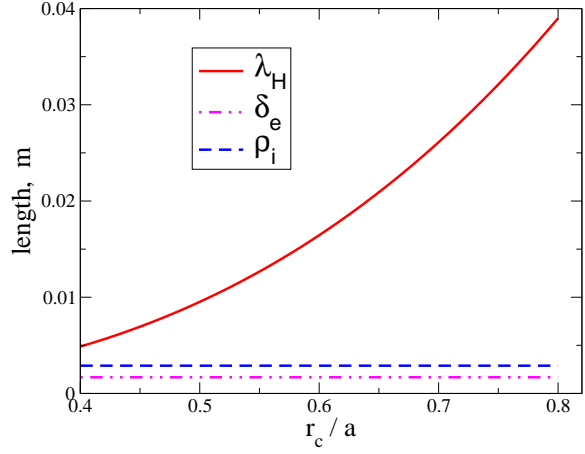
\graph{kink-b10q0.6_qc_scales}{Ideal-MHD inertial-layer width
  $\lambda_H$ as a function of $r_c$ [with $k_{\|}(r_c) = 0$]
  compared to the ion thermal gyroradius $\rho_i$ (computed for $T_i =
  5$~{\rm keV}) and to the electron skin depth $\delta_e$. Screw pinch
  geometry with $R_0 = 5$~{\rm m}, $a = 1$~{\rm m}, $B_0 = 2.5$~{\rm
    T} is considered. For these parameters, the fluid inertial length is larger than
  the kinetic radial scales $\rho_i = 2.8 \times 10^{-3}$~{\rm m} and 
  $\delta_e = 1.7 \times 10^{-3}$~{\rm m}.}

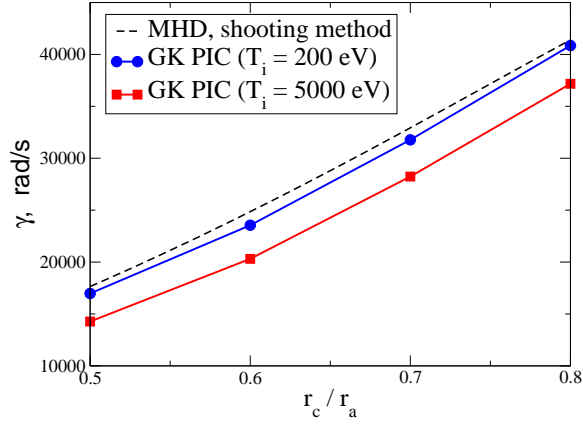
\graphnew{kink1-b10q0.6_qc_ti_g}{Internal kink mode growth rate resulting
  from the gyrokinetic PIC simulations (at two different ion
  temperatures) compared to the results of the ideal-MHD eigenvalue
  calculations (the MHD eigenvalue problem has been numerically solved
  using the shooting method). The agreement between the ideal MHD and
  the gyrokinetic simulations is very good, especially at the smaller ion temperature
  (i.~e.~at the smaller ion gyroradius). One sees that the gyrokinetic kink mode is
  somewhat stabilised when the ion gyroradius increases. This FLR effect
  is absent in the ideal MHD description.} 
% but it is not strong also in
%  the gyrokinetic case, at least for the parameters considered (large
%  $\lambda_H$, i.~e.~strong ideal drive).}

%kink-b10q0.6_qc0.7_ti_g
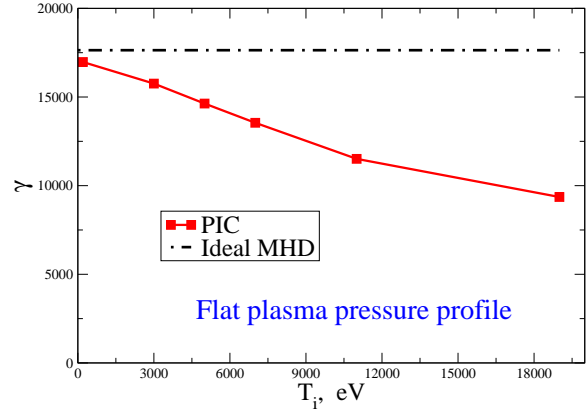
\graphnew{kink1-b10q0.6_qc0.5_ti-g}{Internal kink mode growth rate
  (resulting from the gyrokinetic PIC simulations) as a
  function of the plasma temperature ($T_i = T_e$). The growth rate
  decreases at larger $T_i$ which should be attributed to 
  FLR-stabilization. The rational flux surface is localized at $r_c/a
  = 0.5$. For comparison, the ideal-MHD result is shown (which does not
  depend on the ion temperature).}

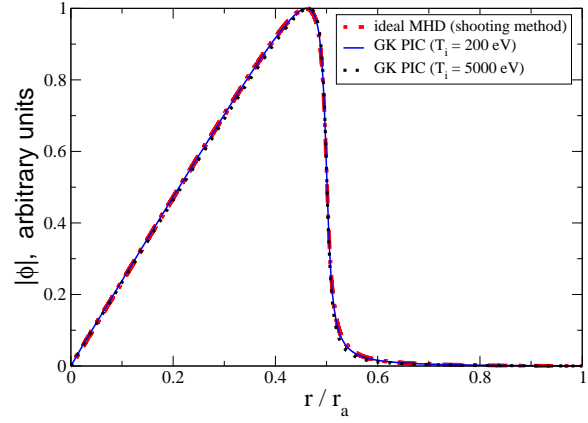
\graphnew{kink1-b10q0.6_qc0.5_ti200_s}{Internal kink mode radial structure in the
  regime with $\lambda_H \gg \rho_i$. The mode changes abruptly at the
  rational flux sruface $r_c = 0.5$ which is a well-known property of
  the intenal kink modes. One sees a perfect agreement between the
  eigenmode structures obtained from the initial-value gyrokinetic PIC
  simulations and solving (numerically) an ideal-MHD eigevalue problem.}

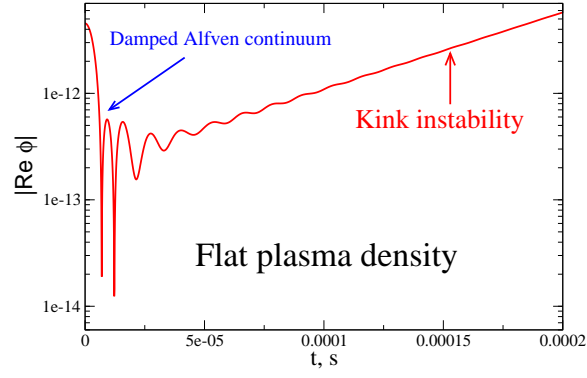
\graph{kink-b10q0.6_qc0.7_t}{Internal kink mode evolution in the
  regime with $\lambda_H \gg \rho_i$. The rational flux sruface is located
  at $r_c = 0.7$. The plasma temperature $T_i = T_e = 5$~{\rm
    keV}. The initial perturbation (a Gaussian) decays in the
  continuum of the shear Alfv\'en waves which are reorganized afterwards 
  into the internal kink eigenmode.} 

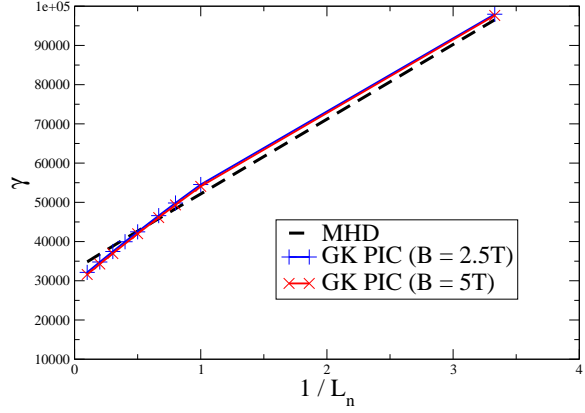
\graphnew{kinkB5-q0.6qc0.7_kn-g}{Pressure gradient effect on the kink
  mode in the MHD regime. Equilibria with ``large'' and ``moderate''
  magnetic fields ($B = 5$~{\rm T} and $B = 2.5$~{\rm T}) have been
  compared. The plasma temperature $T_i = T_e = 3$~{\rm keV}. The
  result does not depend on $B$ (as expected according to MHD
  theory) and the agreement between the gyrokinetic simulations and
  the MHD computation is good.} 

% \graph{kink1-b10q0.6_qc0.7_kn0.4_t}{Diamagnetic effect on the
%   internal kink mode (time evolution) in the MHD regime.}

% \graph{kink1-b10q0.6_qc0.7_kn0.4_s}{Diamagnetic effect on the
%   internal kink mode (radial structure) in the MHD regime. The rational
%   flux surface is   localized at $r_c/a = 0.7$.}

% \graph{kink1-b10q0.6_qc0.7_kn-g}{Diamagnetic effect on the
%   internal kink mode in the MHD regime. Growth rate decreases with the
% density gradient {\it as expected}.}

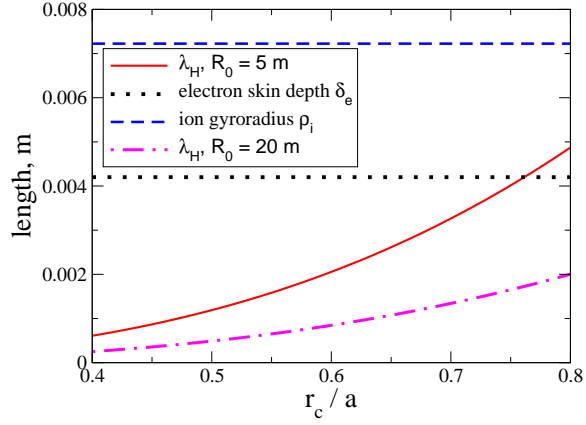
\graph{kink-tp_b1.0q0.6_qc_ti5e3_scales}{Ideal-MHD inertial-layer
  width compared with the ion thermal gyroradius and the electron skin
  depth. Screw pinches with $a = 0.5$~{\rm m},
$B_0 = 1$~{\rm T}, $R_0 = 5$~{\rm m} and $R_0 = 20$~{\rm m} are considered. Here, the
MHD drive is much smaller compared to the case shown in 
Fig.~\ref{kink-b10q0.6_qc_scales} (especially when $R_0 = 20$~{\rm
  m}). As a consequence, the ion gyroradius is much larger than the MHD
inertial length indicating the importance of the FLR effects (and ion
sub-Larmor scales) in the mode evolution (kinetic regime of the $m=1$,
$n = 1$ mode).}  

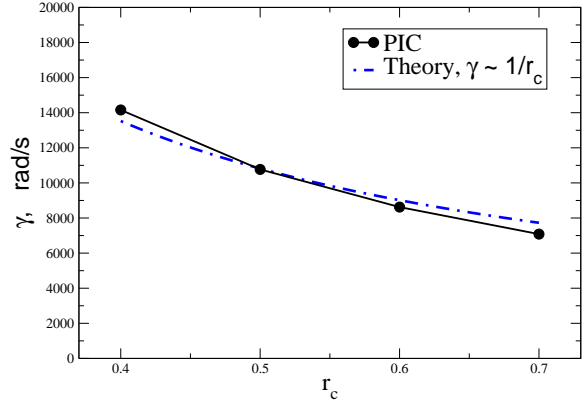
\graphnew{kink-b1.0R20.0qc-gs_ti11e3-g}{Internal kink mode growth rate
  resulting 
  from the gyrokinetic PIC simulations compared to the analytic
  prediction for the collisionless $m=1$ tearing mode \cite{Porcelli91}
  $\gamma \sim (\delta_e \rho_i^2)^{1/3} / r_c$. Here, the parameters
  chosen are $T_i = T_e = 11$~{\rm keV} and $R_0 = 20$~{\rm m}. One
  sees a very good agreement between the theory and simulations.}

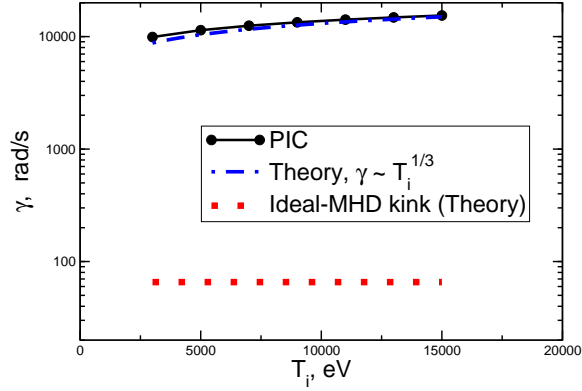
\graphnew{kink_b1.0R20.0qc0.4-gs_ti_g}{Internal kink mode growth rate
  (resulting from the gyrokinetic PIC simulations for $R_0 = 20$~{\rm
    m}) as a function of the plasma temperature ($T_i = T_e$). The growth rate
  increases with the temperature contrary to the ideal-MHD case (see
  Fig.~\ref{kink1-b10q0.6_qc0.5_ti-g}). The rational flux surface is
  localized at $r_c = 0.4$. The agreement of the simulations
  with the theory is very good. For reference, the ideal-MHD growth
  rate corresponding to the parameters chosen is plotted. One sees
  that the growth rate of the kinetic-kink (reconnecting) mode is more than
  two orders of magnitude larger than the growth rate of the marginally-unstable
  ideal-MHD mode.}  

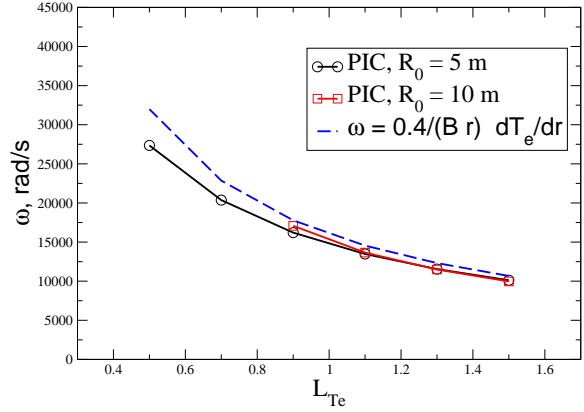
\graphnew{kink-b1.0R5q0.6_qc0.5_kTe-w}{Frequency of the kinetic
  internal kink mode (collisionless $m =1$ tearing mode) as a function
  of the electron temperature gradient. Ion 
  temperature and density are kept constant. For comparison,
  an estimate for the drift-tearing mode frequency $\omega = 0.4 \, 
  \omega_{{\rm T}e}$ \cite{Cowley86} is plotted. One sees that the agreement
  is good. The mode properties for two different aspect ratios (and,
  consequently, $\lambda_H$) are compared (major radius is varied whereas the
  minor radius is kept fixed). One sees that the mode frequency does
  not depend on $\lambda_H$ and, hence, is set exclusively by the diamagnetic
  frequency associated with the electron temperature gradient.}

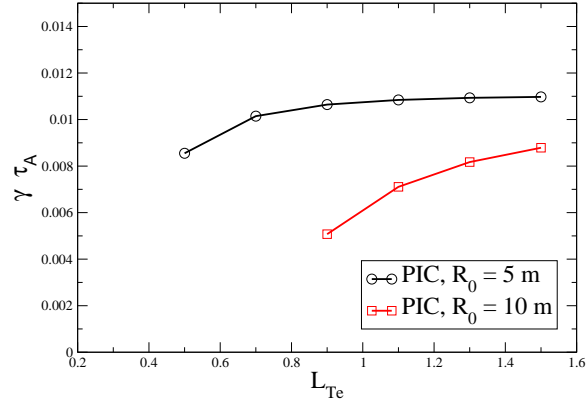
\graphnew{kink-b1.0R5q0.6_qc0.5_kTe-g}{Growth rate of the kinetic
  internal kink mode (collisionless $m =1$ tearing mode) as a function
  of the electron temperature gradient (same parameters as in
  Fig.~\ref{kink-b1.0R5q0.6_qc0.5_kTe-w}). In contrast to the
  frequency, the growth rate is much smaller at larger $R_0$ (note the
  normalization to $\tau_A \propto 1/R_0$.) In this case, the continuum
  damping (set by the mode frequency which is equal for both values of
  $R_0$) corresponds to a larger fraction of the {\it smaller} growth 
  rate, making the stabilization mechanism more effective.} 

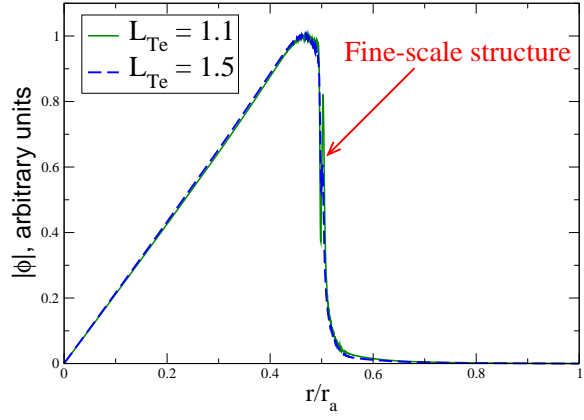
\graphnew{kink-b1.0R10q0.6_qc0.5_kTe_sm}{Kinetic-kink eigenmode 
  ($R_0 = 10$~{\rm m}) compared at different electron temperature
  gradient lengths $L_{{\rm T}e}$. A fine-scale structure develops around the rational flux
  surface [indicating the continuum damping at the position of the resonance
  $\omega = \pm k_{\|}(r_A) v_A$, see
  Fig.~\ref{kink-b1.0R10q0.6_qc0.5_kTe_sm_small} for details]. The fine-scale
  structure is more 
  pronounced (see Fig.~\ref{kink-b1.0R10q0.6_qc0.5_kTe_sm_small}) at
  larger electron temperature gradients (where the mode frequency is
  larger and, consequently, the continuum damping is stronger).} 

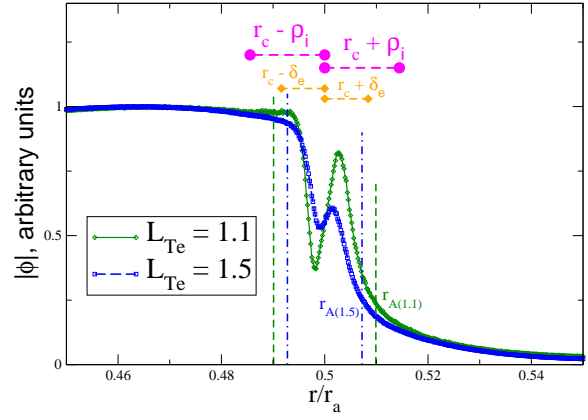
\graphnew{kink-b1.0R10q0.6_qc0.5_kTe_sm_small}{The eigenmodes from
  Fig.~\ref{kink-b1.0R10q0.6_qc0.5_kTe_sm} zoomed around the rational
  flux surface. For comparison, the ion gyroradius and electron skin
  depth scales are plotted. Also, the positions of the shear Alfv\'en
  resonances are indicated (the kinetic Alfv\'en wave resonances would
  be more precise). One sees that the relevant radial scales
  are smaller than the ion gyroradius. Kinetic Alfv\'en waves are 
  excited at the resonant positions causing a stabilization of the
  collisionless $m=1$ drift-tearing mode (through the ``continuum''
  damping mechanism).} 
\end{document}